\documentclass[fleqn,twoside]{article}
\usepackage{espcrc2}

\usepackage{amssymb}

\title{Entropy of a Quantum Oscillator coupled to a Heat Bath and
implications for Quantum Thermodynamics}

\author{{G. W. Ford\address[ford]{Department of Physics, University of
Michigan, Ann Arbor, Michigan 48109-1120}}, R. F.
O'Connell\address{Department of Physics and Astronomy, Louisiana State
University, Baton Rouge, Louisiana 70803-4001}}

\begin{document}

\begin{abstract}
The free energy of a quantum oscillator in an arbitrary heat bath at
temperature $T$ is given by a "remarkable formula" which involves only a
single integral.  This leads to a corresponding simple result for the
entropy.    The low temperature
limit is examined in detail and we obtain explicit results both for the case of an
Ohmic heat bath and a radiation heat bath.  More general heat bath models are also
examined.  This enables us to
determine the entropy at zero temperature in order to check the third law of
thermodynamics in the quantum regime.
\vspace{1pc}
\end{abstract}

\maketitle

\section{Introduction}

The widespread interest in recent years in (a) mescoscopic systems
\cite{grinstein,ferry,datta,imry,bergmann,lee,chakravasty} and (b)
fundamental quantum physics and quantum computation
\cite{bennett,quantuminfo,haroche,zeilinger,giulini,myatt,ford1,ford2,ford3}
has highlighted the critical role which dissipative environments play in
such studies.  This has led to a critical examination of many results that
were derived for macroscopic systems.  In particular, there has been
considerable interest in the area of quantum and mesoscopic
thermodynamics, the subject of this conference.  In particular, in some
instances questions have been raised about the validity of the
fundamental laws of thermodynamics.  Whereas many interesting new facets of
old results have emerged, it is important to exercise caution before
questioning the validity of fundamental laws (especially the three laws
of thermodynamics), since many subtle issues arise.

Here, we examine the third law of thermodynamics in the quantum regime by
calculating the entropy $S$ for a quantum oscillator in an arbitrary heat
bath at temperature $T$ and checking to see if it vanishes as
$T\rightarrow 0$, in conformity with Nernst's law \cite{landau}.

The first question which arises is how to calculate $S$.  The von Neumann
result for $S$ involves $\rho\log\rho$, where $\rho$ is the density
matrix for the whole system, a non-trivial quantity to calculate.  As a
result, one's first thought is perhaps to make use of the Wigner
distribution function $W$ corresponding to $\rho$.  This has led some
authors to simply replace $\rho$ with $W$ in the von Neumann result but,
unfortunately, this is not correct because the $W$ corresponding to $\log\rho$
is not the
$\log$ of the $W$ corresponding to $\rho$ and, as a consequence, one is led to
misleading conclusions.  A way out of this impasse is to use the method
which we introduced in 1985, in collaboration with J.T. Lewis \cite{ford4}, in
order to calculate the free energy $F$.  Then, using a familiar thermodynamic
relation, the result for $S$ readily follows.  In addition, the result for the
total energy
$U$ follows in a similar manner.  In Section 2, we review this method and
write down the results for $F$, $S$, and $U$ in the case of a quantum
oscillator in an arbitrary heat bath at temperature $T$.  All of these
results involve just a single integral.  In section 3, we evaluate the
relevant integral for the case of an Ohmic heat bath and for an arbitrary
$T$.  This enables us to show that $S\rightarrow 0$ as $T\rightarrow 0$,
in conformity with Nernst's law.  A similar result is obtained in Section
4 for the case of a blackbody radiation heat bath and in Section 5 for an
arbitrary heat bath.  Our conclusions are given in Section 6.

\section{Fundamentals}

A very general model for the motion of a quantum particle in an arbitrary
heat bath is the independent-oscillator model \cite{ford1,ford4}, which
is described by the Hamiltonian
\begin{eqnarray}
H&=&\frac{p^{2}}{2m}+V(x)
\nonumber
\\ 
&&{}+\sum_{j}\left(\frac{p^{2}_{j}}{2m_{j}}+\frac{1}{2}m_{j}\omega^{2}_{j}
\left(q_{j}-x\right)^{2}\right).
\label{eq1}
\end{eqnarray}
Here $m$ is the mass of the quantum particle while $m_{j}$ and
$\omega_{j}$ refer to the mass and frequency of heat-bath oscillator
$j$.  In addition, $x$ and $p$ are the coordinate and momentum operators
for the quantum particle and $q_{j}$ and $p_{j}$ are the corresponding
quantities for the heat-bath oscillators.  The infinity of choices for
the $m_{j}$ and $\omega_{j}$ give this model its great generality.  Moreover, as
emphasized in \cite{ford1}, the most general coupling of a quantum particle to a
linear passive heat bath is equivalent with an independent oscillator model, which
is described by the Hamiltonian given in (\ref{eq1}).

Use of the Heisenberg equations of motion leads to the 
quantum Langevin equation
\begin{eqnarray}
m\ddot{x}&+&\int^{t}_{-\infty}{\textnormal{d}}t^{\prime}\mu
(t-t^{\prime})\dot{x}(t^{\prime})
+V^{\prime}(x) \nonumber \\
&&=F(t) \label{eq2}
\end{eqnarray}
where $V^{\prime}(x)={\textnormal{d}}V(x)/{\textnormal{d}}x$ is the
negative of the time-independent external force and $\mu(t)$ is the
so-called memory function.  $F(t)$ is the random (fluctuation or noise)
operator force with mean $\langle F(t)\rangle =0$.  The quantities $\mu
(t)$ and
$F(t)$ describe the properties of the heat bath.

In the particular case of an oscillator potential
\begin{equation}
V(x)=\frac{1}{2}Kx^{2}=\frac{1}{2}m\omega^{2}_{0}x^{2}. \label{eq3}
\end{equation}
Substituting (\ref{eq3}) into (\ref{eq2}) enables us to obtain the
explicit solution
\begin{equation}
\tilde{x}(\omega )=\alpha (\omega )\tilde{F}(\omega ), \label{eq4}
\end{equation}
where the superposed tilde is used to denote the Fourier transform. 
Thus, $\tilde{x}(\omega )$ is the Fourier transform of the operator
$x(t)$:
\begin{equation}
\tilde{x}(\omega )=\int^{\infty}_{-\infty}{\textnormal{d}}tx(t)e^{i\omega
t},
\label{eq5}
\end{equation}
and similarly for $\tilde{F}(\omega )$. Here
$\alpha (z)$ is the familiar response function (generalized
susceptibility)
\begin{equation}
\alpha (z)=\frac{1}{-mz^{2}-iz\tilde{\mu}(z)+K}. \label{eq6}
\end{equation}
and $\tilde{\mu}(z)$ is the Fourier transform of the memory function:
\begin{equation}
\tilde{\mu}(z)=\int^{\infty}_{0}{\textnormal{d}}t\mu (t)e^{izt}.
\label{eq7}
\end{equation}

We have now all the tools at our disposal necessary to obtain
thermodynamic qualities for the heat bath.  Our main task will be the
calculation of the free energy $F$, which is a thermodynamic potential from which
other thermodynamic functions can be obtained by differentiation.  The entropy is
the only one of interest here and is given by the relation  
\begin{equation}
S=-\left(\frac{\partial F}{\partial T}\right)_{V}. \label{eq8}
\end{equation}

The system of an oscillator coupled to a heat bath in thermal equilibrium
at temperature $T$ has a well-defined free energy.  The free energy
ascribed to the oscillator, $F(T)$, is given by the free energy of the
system minus the free energy of the heat bath in the absence of the
oscillator.  This calculation was carried out by two different methods
\cite{ford4,ford5} leading to the "remarkable formula"
\begin{equation}
F(T)=\frac{1}{\pi}\int^{\infty}_{0}d\omega f(\omega
,T){\textnormal{Im}}\left(\frac{d~\log\alpha (\omega +i0^{+})}{d\omega}\right),
\label{eq9}
\end{equation}
where $f(\omega ,T)$ is the free energy of a single oscillator of
frequency $\omega$, given by
\begin{equation}
f(\omega ,T)=kT\log [1-\exp\left(-\hbar\omega /kT\right)]. \label{eq10}
\end{equation}
where the zero-point contribution $(\hbar\omega /2)$ has been omitted. 
Thus, all that remains is to specify $\tilde{\mu}(z)$ which characterizes
the heat bath.  In the remaining sections, we consider various heat bath models.  In
the low temperature case $(kT<<\hbar\omega_{0})$, explicit results may be obtained
by noting that $f(\omega ,T)$ vanishes exponentially for large $(\hbar\omega /kT)$. 
Hence, the integrand in (\ref{eq9}) is confined to small $(\hbar\omega /kT)$ so
that the factor multiplying $f(\omega ,T)$ can be expanded in powers of $\omega$.

\section{Ohmic heat bath}

This is an oft-studied model for which
\begin{equation}
\tilde{\mu}(\omega )=m\gamma, \label{eq11}
\end{equation}
where $\gamma$ is a constant.  Thus
\begin{equation}
\alpha(\omega
)\equiv\left[m\left(\omega^{2}_{0}-\omega^{2}-i\omega\gamma\right)\right]
^{-1} \label{eq12}
\end{equation}
is the familiar phenomenological Drude-Lorentz model result.  Hence,
using (\ref{eq12}) and (\ref{eq10}) in (\ref{eq9}), we obtain
\begin{eqnarray}
F(T)&=\frac{kT}{\pi}\int^{\infty}_{0}d\omega ~\log\left[1-\exp
\left(-\hbar\omega/kT\right)\right]\nonumber \\
&\times~\frac{\gamma\left(\omega^{2}+\omega^{2}_{0}\right)}{\left[
\left(\omega^{2}-\omega^{2}_{0}\right)^{2}+\gamma^{2}\omega^{2}\right]}.
\label{eq13}
\end{eqnarray}
Hence, in the low temperature case,
\begin{eqnarray}
{\textnormal{Im}}&\left\{\frac{d\log\alpha (\omega
)}{d\omega}\right\}=\frac{\gamma\left(\omega^{2}+\omega^{2}_{0}\right)}
{\left(\omega^{2}_{0}-\omega^{2}\right)^{2}+\gamma^{2}\omega^{2}}
\nonumber \\
&\rightarrow\frac{\gamma}{\omega^{2}_{0}}. \label{eq14}
\end{eqnarray}
Substituting into (\ref{eq13}) and changing the variable of integration to
\begin{equation}
y=\frac{\hbar\omega}{kT}, \label{eq15}
\end{equation}
we obtain
\begin{equation}
F(T)=\frac{\gamma
(kT)^{2}}{\pi\hbar\omega^{2}_{0}}\int^{\infty}_{0}dy\log\left(1-e^{-y}
\right). \label{eq16}
\end{equation}
The following integral is relevant (now and later)
\begin{equation}
\int^{\infty}_{0}dyy^{z}\log (1-e^{-y})=-\Gamma (z+1)\zeta (z+2),
\label{eq17}
\end{equation}
where $\zeta$ is the Riemann zeta-function,
\begin{equation}
\zeta (z)=\sum^{\infty}_{n=1}~\frac{1}{n^{z}}. \label{eq18}
\end{equation}
If $z$ is an even integer $\zeta (z)$ is related to the Bernoulli numbers, $\zeta
(2)=\pi^{2}B_{1}=\pi^{2}/6,~\zeta (4)=\pi^{4}B_{2}/3=\pi^{4}/90$, etc.  But in
Section 5 other values appear. Thus, in particular,
\begin{equation}
\int^{\infty}_{0}dy\log\left(1-e^{-y}\right)=-\frac{\pi^{2}}{6},
\label{eq19}
\end{equation}
\begin{equation}
\int^{\infty}_{0}dyy^{2}\log\left(1-e^{-y}\right)=-\frac{\pi^{4}}{45}.
\label{eq20}
\end{equation}
Hence
\begin{equation}
F(T)=-\frac{\pi}{6}\hbar\gamma\left(\frac{kT}{\hbar\omega_{0}}\right)^{2}.
\label{eq21}
\end{equation}
Also
\begin{equation}
S(T)=-\left(\frac{\partial F}{\partial T}\right)_{V}=\frac{\pi}{3}\gamma\frac
{k^{2}T}{\hbar\omega^{2}_{0}}. \label{eq22}
\end{equation}

We note that $S(T)\rightarrow 0$ as $T\rightarrow 0$, in conformity with
the third law of thermodynamics.

\section{Blackbody radiation heat bath}

In this case, we obtained \cite{ford6}
\begin{equation}
\alpha(\omega )=(1-i\omega\tau_{e})\alpha_{D}(\omega ), \label{eq23}
\end{equation}
where $\alpha_{D}(\omega )$ is the Drude result with $\gamma$ replaced by
\begin{equation}
\gamma_{e}=\omega^{2}_{0}\tau_{e}, \label{eq24}
\end{equation}
and
\begin{equation}
\tau_{e} = \frac{2e^{2}}{3mc^{3}}
= 6\times 10^{-24}s. \label{eq25}
\end{equation}
Thus, proceeding as in the Ohmic case and once again letting $\omega\rightarrow 0$
in order to calculate the results for small $T$, we obtain \cite{ford7}
\begin{eqnarray}
{\textnormal{Im}}\left\{\frac{d\log\alpha(\omega
)}{d\omega}\right\}{\!\!\!\!\!\!}&=&{\!\!\!}
\frac{\gamma_{e}
\left(\omega^{2}+\omega^{2}_{0}\right)}{\left(\omega^{2}_{0}-\omega^{2}\right)^{2}+
\gamma^{2}_{e}\omega^{2}}-\frac{\gamma_{e}}{1+\omega^{2}\tau^{2}_{e}}
\nonumber \\
&&\rightarrow\frac{3\gamma_{e}}{\omega^{4}_{0}}\omega^{2} . \label{eq26}
\end{eqnarray}
It follows that 
\begin{eqnarray}
F(T) &=& \frac{\gamma_{e}
(kT)^{2}}{\pi\hbar\omega^{2}_{0}}3\left(\frac{kT}{\hbar\omega_{0}}\right)^{2}
\int^{\infty}_{0}dyy^{2}\log(1-e^{-y}) \nonumber \\
&=&
-\frac{\pi^{3}}{15}\hbar\gamma_{e}\left(\frac{kT}{\hbar\omega_{0}}
\right)^{4}, \label{eq27}
\end{eqnarray}
from which it follows that
\begin{equation}
S=\frac{4\pi^{2}}{15}\hbar\gamma_{e}\frac{k^{4}T^{3}}{(\hbar\omega_{0})^{4}}.
\label{eq28}
\end{equation}
Once again, we see that $S\rightarrow 0$ as $T\rightarrow 0$, in
agreement with Nernst's law.  In this case, $S\rightarrow 0$ faster than in the
Ohmic case, as a result of the fact that the right-side of (\ref{eq25}) has a factor
$\omega^{2}$ whereas the corresponding result on the right-side of (\ref{eq14}) is
independent of $\omega$.

\section{Arbitrary heat bath}

From (\ref{eq3}) and (\ref{eq6}), we obtain
\begin{equation}
{\textnormal{Im}}~\alpha (\omega )=\omega\vert\alpha(\omega
)\vert^{2}~{\textnormal{Re}}~\tilde{\mu}(\omega ), \label{eq29}
\end{equation}
\begin{eqnarray}
{\textnormal{Re}}~\alpha (\omega
)&{\!\!\!\!\!\!\!\!\!\!\!\!\!\!\!\!\!\!\!}=\left\{-m\left(\omega^{2}-\omega^{2}_{0}\right)\right.
\nonumber \\
&\left.-\omega~{\textnormal{Im}}~
\tilde{\mu}(\omega )\right\}\vert\alpha(\omega )\vert^{2}. \label{eq30} 
\end{eqnarray}
Using these results in (\ref{eq9}) leads to
\begin{eqnarray}
&F(T)=\frac{1}{\pi}\int^{\infty}_{0}d\omega f(\omega
,T)\vert\alpha\vert^{2}\left\{m\left(\omega^{2}+\omega^{2}_{0}\right)\right.
\nonumber \\ 
&{\textnormal{Re}}~\tilde{\mu}(\omega )
\left.-\omega^{2}{\textnormal{Re}}~\tilde{\mu}(\omega )\frac{d}{d\omega}
~{\textnormal{Im}}~\tilde{\mu}(\omega )\right. \nonumber \\
&+\left.\omega\left(-m\omega^{2}+m\omega^{2}_{0}+\omega{\textnormal{Im}}~
\tilde{\mu}\right)\frac{d}{d\omega}{\textnormal{Re}}~\tilde{\mu}(\omega )
\right\}. \label{eq31}
\end{eqnarray}
Now we make use of the fact that $\tilde{\mu}(z)$ must be a positive real function
\cite{ford1} and hence the boundary value of
$\tilde{\mu}(z)$ on the real axis has everywhere a positive real part i.e.
\begin{equation}
{\textnormal{Re}}\left[\tilde{\mu}\left(\omega +i0^{+}\right)\right]\ge0,~~-\infty
< \omega <\infty. \label{eq32}
\end{equation}
Thus, in the neighborhood of the origin, a very general class of models are
incorporated by writing
\begin{equation}
\tilde{\mu}(z)=mb^{1-\alpha}(-iz)^{\alpha},~~-1\leq\alpha \leq 1, \label{eq33}
\end{equation}
where $b$ is a constant with the dimensions of frequency.  It is easy to verify that
this is a positive real function if and only if $\alpha$ is within the indicated
range [we choose the branch $-\pi <\theta <\pi$ where $\theta$ is $arg(z)$].  Hence
\begin{eqnarray}
&{\textnormal{Re}}~\tilde{\mu}(\omega
)\sim\omega^{\alpha}\cos\left\{\alpha\left(\theta -\pi /2\right)\right\} \nonumber
\\
&{\textnormal{Im}}~\tilde{\mu}(\omega
)\sim\omega^{\alpha}\sin\left\{\alpha\left(\theta -\pi /2\right)\right\}.
\label{eq34}
\end{eqnarray}
As it turns out, the case $\alpha =-1$ requires special treatment so we will
consider this first, obtaining $\tilde{\mu}=imb^{2}/z$, so that
\begin{equation}
\alpha (\omega)=\frac{1}{-m\omega^{2}+m(b^{2}+\omega^{2}_{0})}. \label{eq35}
\end{equation}
Thus, this corresponds to a shift in the force constant $K$ with no dissipation.  In
the absence of dissipation
\begin{eqnarray}
{\textnormal{Im}}\left\{\frac{d\log\alpha (\omega
+i0^{+})}{d\omega}\right\}\!\!\!\!\!&=\pi\left[\delta(\omega -\omega_{0})\right.
\nonumber \\
&{}\left.+\delta (\omega +\omega_{0})\right],
\label{eq36}
\end{eqnarray}
so
\begin{equation}
F(T)=f(\omega_{0},T)\cong -kTe^{-\hbar\omega /kT}, \label{eq37}
\end{equation}
from which it is clear that $S(T)\rightarrow 0$ as $T\rightarrow 0$. We now return
to the more general case $-1<\alpha \leq 1$ and again consider the low-temperature
limit. Thus, as
$\omega\rightarrow 0$, the
\{
\} term in (\ref{eq30}) reduces to
\begin{equation}
\{~\}\rightarrow m\omega^{2}_{0}(1+\alpha ){\textnormal{Re}}~\tilde{\mu},
\label{eq38}
\end{equation}
and, in addition, $\vert\alpha (\omega )\vert^{2}\rightarrow
\left(m\omega^{2}_{0}\right)^{-2}$.  It follows that
\begin{equation}
F(T)\rightarrow \!\frac{kT}{\pi
m\omega^{2}_{0}}b^{(1-\alpha)}(1+\alpha)\cos\left\{\alpha
\left(\theta-\frac{\pi}{2}\right)\right\}I(\omega), \label{eq39}
\end{equation}
where
\begin{eqnarray}
I(\omega )&=
\int^{\infty}_{0}d\omega\omega^{\alpha}\log\left[1-\exp\left(-\hbar\omega
/kT\right)\right] \nonumber \\
&=\left(\frac{kT}{\hbar}\right)^{\alpha+1}\int^{\infty}_{0}dyy^{\alpha}
\log\left(1-e^{-y}\right). \label{eq40}
\end{eqnarray}
It follows from (\ref{eq39}), (\ref{eq40}) and (\ref{eq8}) that
\begin{eqnarray}
S(T)&\!\!\!\!=\frac{1}{\pi m\omega^{2}_{0}}b^{(1-\alpha)}(\alpha
+1)(\alpha+2)\frac{k^{\alpha +2}}{\hbar^{\alpha +1}} \nonumber \\
&\times T^{\alpha +1}\Gamma (\alpha +1)\xi (\alpha +2). \label{eq41}
\end{eqnarray}
As a check, we note that for $\alpha =0$ and $b=\gamma$, this result reduces to the
Ohmic result given in (\ref{eq22}).  Hence, since $(\alpha +1)$ can never be
negative, we conclude that
\begin{equation}
S(T)\sim(kT)^{\alpha +1}\longrightarrow 0~~ as~T\rightarrow 0. \label{eq42}
\end{equation}

\section{Conclusions}

For the case of a quantum oscillator coupled to an arbitrary heat bath,
we found that Nernst's third law of thermodynamics is still valid.

\section{Acknowledgements}

One of us (RFOC) is pleased to acknowledge an enlightening conversation with
W.P. Schleich on the question of the feasability of using the Wigner
distribution in order to calculate $S$.


\begin{thebibliography}{00}

\bibitem{grinstein} G. Grinstein, G. Mazenko (Eds.), Directions in
Condensed Matter Physics, World Scientific Press, Singapore, 1986.

\bibitem{ferry} D.K. Ferry, J.R. Barker, C. Jacobini (Eds.), Granular
Nanoelectronics Plenum Press, New York, 1990.

\bibitem{datta} S. Datta, Electronic Transport in Mesoscopic Systems,
Cambridge University Press, Cambridge, 1997.

\bibitem{imry} Y. Imry, Introduction to Mesoscopic Physics, Oxford
University Press, Oxford, 1997.

\bibitem{bergmann} G. Bergmann, Phys. Rep. 107 (1984) 1.

\bibitem{lee} P.A. Lee, T.V. Ramakrishnan, Rev. Mod. Phys. 57 (1985) 287.

\bibitem{chakravasty} S. Chakravasty, A. Schmid, Phys. Rep. 140 (1986)
193.

\bibitem{bennett} C.H. Bennett, Phys. Today 48 (10) (1995) 24.

\bibitem{quantuminfo} Quantum Information [Special Issue of Phys. World
11(3) (1998)].

\bibitem{haroche} S. Haroche, Phys. Today 51 (7) (1998) 36.

\bibitem{zeilinger} A. Zeilinger, Sci. Am. 282 (4) (2000) 32.

\bibitem{giulini} D. Giulini, E. Joos, C. Kiefer, J. Kupischg, I.-O.
Stamatescu, H.D. Zeh, Decoherence and the Appearance of a Classical World
in Quantum Theory, Springer, New York, 1996.

\bibitem{myatt} C.J. Myatt, B.E. King, Q.A. Turchette, C.A. Sackett, D.
Kielpinski, W.M. Itano, C. Monroe, D.J. Wineland, Nature (London) 403
(2000) 269.

\bibitem{ford1} G.W. Ford, J.T. Lewis, R.F. O'Connell, Phys. Rev. A 37
(1988) 4419.

\bibitem{ford2} G.W. Ford, J.T. Lewis, R.F. O'Connell, Phys. Rev. A 64
(2001) 032101.

\bibitem{ford3} G.W. Ford, R.F. O'Connell, Phys. Rev. D 64 (2001) 105020.

\bibitem{landau} L.D. Landau, E.M. Lifshitz, L.P.; Pitaevskii,
Statistical Physics, 3rd Edition, Part 1, Pergamon Press, New York, 1980.

\bibitem{ford4} G.W. Ford, J.T. Lewis, R.F. O'Connell, Phys. Rev. Lett.
55 (1985) 2273.

\bibitem{ford5} G.W. Ford, J.T. Lewis, R.F. O'Connell, Ann. Phys. (NY) 185
(1988) 270.

\bibitem{ford6} G.W. Ford, J.T. Lewis, R.F. O'Connell, Phys. Rev. A 36
(1987) 1466.

\bibitem{ford7} G.W. Ford, J.T. Lewis, R.F. O'Connell, J. Phys. B 20
(1987) 899.

\end{thebibliography}
\end{document}